\pdfoutput=1

\PassOptionsToPackage{table}{xcolor}
\documentclass[sigconf,authorversion]{acmart}
\AtBeginDocument{%
  }

\setcopyright{acmlicensed}
\copyrightyear{2025}
\acmYear{2025}
\setcopyright{acmlicensed}\acmConference[ICPE Companion '25]{Companion of the 16th ACM/SPEC International Conference on Performance Engineering}{May 5--9, 2025}{Toronto, ON, Canada}
\acmBooktitle{Companion of the 16th ACM/SPEC International Conference on Performance Engineering (ICPE Companion '25), May 5--9, 2025, Toronto, ON, Canada}
\acmDOI{10.1145/3680256.3721306}
\acmISBN{979-8-4007-1130-5/2025/05}

\usepackage{siunitx} 
\usepackage[T1]{fontenc}
\usepackage[utf8]{inputenc}
\usepackage{pmboxdraw}
\usepackage{caption}
\usepackage{subcaption}
\usepackage{xspace}
\usepackage{csquotes}
\usepackage{fancyvrb}
\usepackage{makecell}
\usepackage{multirow}

\newcommand{\thename}{cfdSCOPE\xspace}
\newcommand{\acronymexplanation}{short for \textit{\textbf{c}omputational \textbf{f}luid \textbf{d}ynamics \textbf{S}imulation \textbf{C}ode for \textbf{O}ptimization and \textbf{P}erformance \textbf{E}ngineering}}

\begin{document}

\title{\thename: A Fluid-Dynamics Proxy App for Teaching Performance Engineering}


\newcommand{\lbcluster}{\ifx\myanonymous\undefined Lichtenberg high-performance computer at TU Darmstadt\else Anonymous HPC cluster\fi}
\newcommand{\repo}{\ifx\myanonymous\undefined \url{https://github.com/tudasc/cfdSCOPE}\else \url{https://anonymous.4open.science/r/cfdSCOPE-5224}\fi}

\ifx\myanonymous\undefined

\author{Peter Arzt}
\email{peter.arzt@tu-darmstadt.de}
\orcid{0000-0001-6937-1158}
\affiliation{%
  \institution{Technical University of Darmstadt}
  \city{Darmstadt}
  \state{Hesse}
  \country{Germany}
}

\author{Sebastian Kreutzer}
\email{sebastian.kreutzer@tu-darmstadt.de}
\orcid{0000-0002-1641-4342}
\affiliation{%
  \institution{Technical University of Darmstadt}
  \city{Darmstadt}
  \state{Hesse}
  \country{Germany}
}

\author{Tim Jammer}
\email{tim.jammer@tu-darmstadt.de}
\orcid{0000-0003-3735-9677}
\affiliation{%
  \institution{Technical University of Darmstadt}
  \city{Darmstadt}
  \state{Hesse}
  \country{Germany}
}

\author{Christian Bischof}
\email{christian.bischof@tu-darmstadt.de}
\orcid{0000-0003-2711-3032}
\affiliation{%
  \institution{Technical University of Darmstadt}
  \city{Darmstadt}
  \state{Hesse}
  \country{Germany}
}

\else

\author{First Author}
\email{author1@example.com}
\orcid{xxxx-xxxx-xxxx-xxxx}
\affiliation{%
  \institution{Anonymous institution}
  \city{City}
  \state{State}
  \country{Country}
}

\author{Second Author}
\email{author2@example.com}
\orcid{xxxx-xxxx-xxxx-xxxx}
\affiliation{%
  \institution{Anonymous institution}
  \city{City}
  \state{State}
  \country{Country}
}

\author{Third Author}
\email{author3@example.com}
\orcid{xxxx-xxxx-xxxx-xxxx}
\affiliation{%
  \institution{Anonymous institution}
  \city{City}
  \state{State}
  \country{Country}
}

\author{Fourth Author}
\email{author4@example.com}
\orcid{xxxx-xxxx-xxxx-xxxx}
\affiliation{%
  \institution{Anonymous institution}
  \city{City}
  \state{State}
  \country{Country}
}

\fi


\begin{abstract}
  Teaching performance engineering in high-performance computing (HPC) requires example codes that demonstrate bottlenecks and enable hands-on optimization.
  However, existing HPC applications and proxy apps often lack the balance of simplicity, transparency, and optimization potential needed for effective teaching.  
  To address this, we developed \thename, a compact, open-source computational fluid dynamics (CFD) proxy app specifically designed for educational purposes. \thename simulates flow in a 3D volume using sparse linear algebra, a common HPC workload, and comprises fewer than 1,100 lines of code.
  Its minimal dependencies and transparent design ensure students can fully control and optimize performance-critical aspects, while its naive OpenMP parallelization provides significant optimization opportunities, thus making it an ideal tool for teaching performance engineering.
\end{abstract}




\maketitle

\ifx\abstractonly\undefined

\vspace{3cm}
\section{Introduction}
\label{sec:intro}
Teaching students performance engineering is essential for the development of efficient, scalable, and sustainable applications, especially in the area of high-performance computing (HPC).
Understanding how to identify and address bottlenecks ensures that students can improve resource utilization and adapt applications to evolving hardware architectures. 
When teaching performance analysis and optimization, educators often provide students with an example code to showcase recurring performance problems and allow for student experimentation.
The selection of such an example code is mostly driven by these requirements:
\begin{enumerate}
    \item
    The code should include \textbf{performance-relevant structures}, such as stencil or linear algebra computations typical of HPC.
    Furthermore, as \textbf{parallel computing} is essential to exploit any modern hardware architecture, the implemented algorithms should offer opportunities for parallelization or already be expressed using an established programming model for parallel computing.

    \item
    The code should be distributed under an \textbf{open-source license} to allow code redistribution to students without risking license infringements.

    \item
    The application should have \textbf{manageable size and complexity}, enabling students to become familiar and productive within the limited duration of a course.

    \item
    A \textbf{minimal number of dependencies} is preferable to give students full control over its performance characteristics.
    Applications that offload significant workloads to external libraries (e.g., linear algebra solvers) are less ideal, as they obscure performance-critical computations.
    Additionally, since we want to focus students' time on improving the code, we want to keep the effort for compiling and linking as simple as possible.
    
    \item
    Most importantly, the application should exhibit significant \textbf{optimization potential}, particularly in the areas covered by the course.
    To maximize learning outcomes, students should have opportunities to apply common optimization techniques through a learning-by-doing approach, rather than merely analyzing pre-optimized code.
    Furthermore, already highly optimized codes often poses a steep learning curve due to low readability, especially when using advanced programming concepts like intrinsics or inline assembly.
\end{enumerate}

A myriad of HPC codes exists, but consideration of the above requirements severely narrows the range of suitable candidates for use in teaching:
For example, HPC educators could choose to use real-world, production HPC applications which they are familiar with from past or present work.
While such a choice offers the advantage of practical relevance and it can be assumed that some parts of such applications have potential for optimization (especially when considering modern hardware generations), production HPC codes usually have a large and complex code base which can appear daunting to HPC beginners.
Even for students experienced in navigating complex software projects, the effort required to set up, compile and execute most production HPC application often presents a steep entry barrier as substantial time is required to get familiar with the code structure.

To identify example codes of more manageable size, one could explore smaller, well-established HPC benchmarks.
Proxy apps such as LLNL's LULESH~\cite{LULESH} or AMG2013~\cite{AMG} typically consist of only a few thousand lines of code and are designed to encapsulate the performance characteristics and key features of larger applications in a condensed format.
This compactness can make proxy apps appealing for teaching purposes.
However, since these proxy apps are intended as low-threshold test beds for performance experimentation, they are often accompanied by detailed descriptions of optimization efforts in the literature or are already highly optimized.
Consequently, they may offer limited opportunities for further optimization, particularly at the beginner or intermediate levels typically targeted in performance engineering courses.

Another option is to extract relevant sections from a large, productive HPC code that exhibit optimization potential, creating a custom mini app.
While this method has the potential to fulfill the teaching requirements (assuming the original code’s license permits modification and redistribution), the extraction process is non-trivial and time-consuming.
It requires a thorough understanding of data flow and often domain-specific expertise~\cite{lehr_mini_app_extraction}.
Although automatic extraction methods exist, they are currently of limited utility as they are restricted to applications composed of a single translation unit~\cite{lehr_mini_app_extraction} or produce output in a lower-level representation (LLVM~IR)~\cite{castro_cere}, which is unsuitable for this use case. 

\begin{figure*}
    \centering
    \begin{minipage}{.93\linewidth}
    \begin{subfigure}{0.24\linewidth}
        \centering
        \includegraphics[width=\linewidth]{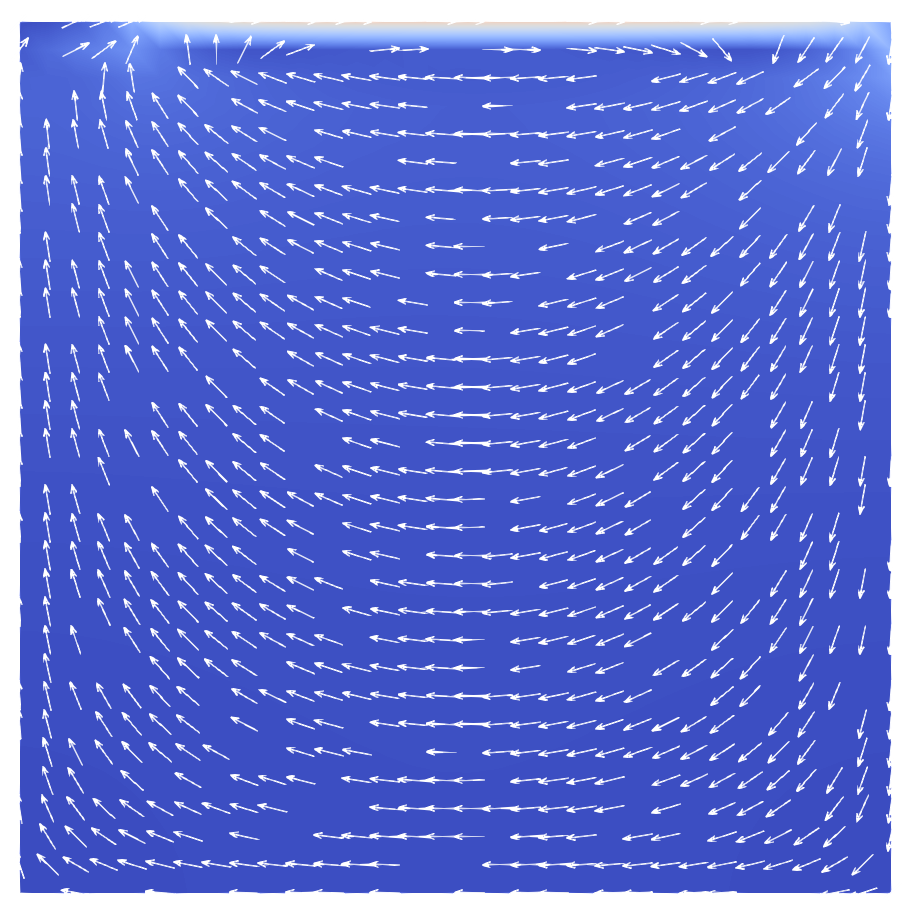}
    \end{subfigure}
    \begin{subfigure}{0.24\linewidth}
        \centering
        \includegraphics[width=\linewidth]{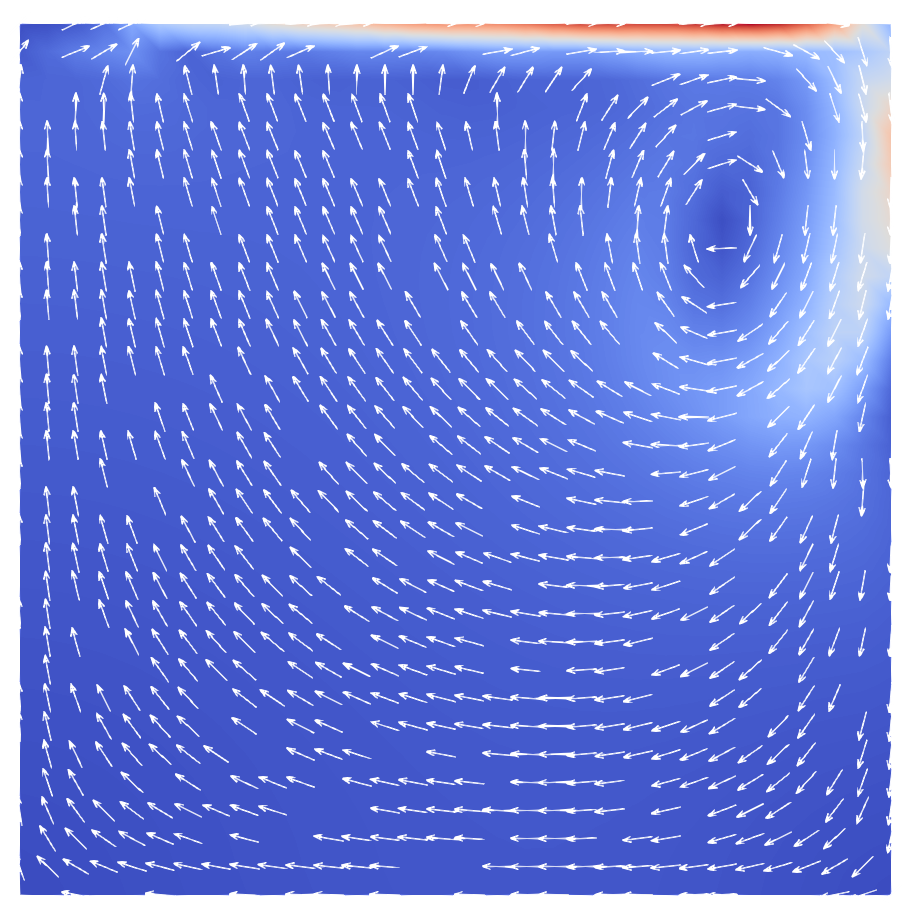}
    \end{subfigure}
    \begin{subfigure}{0.24\linewidth}
        \centering
        \includegraphics[width=\linewidth]{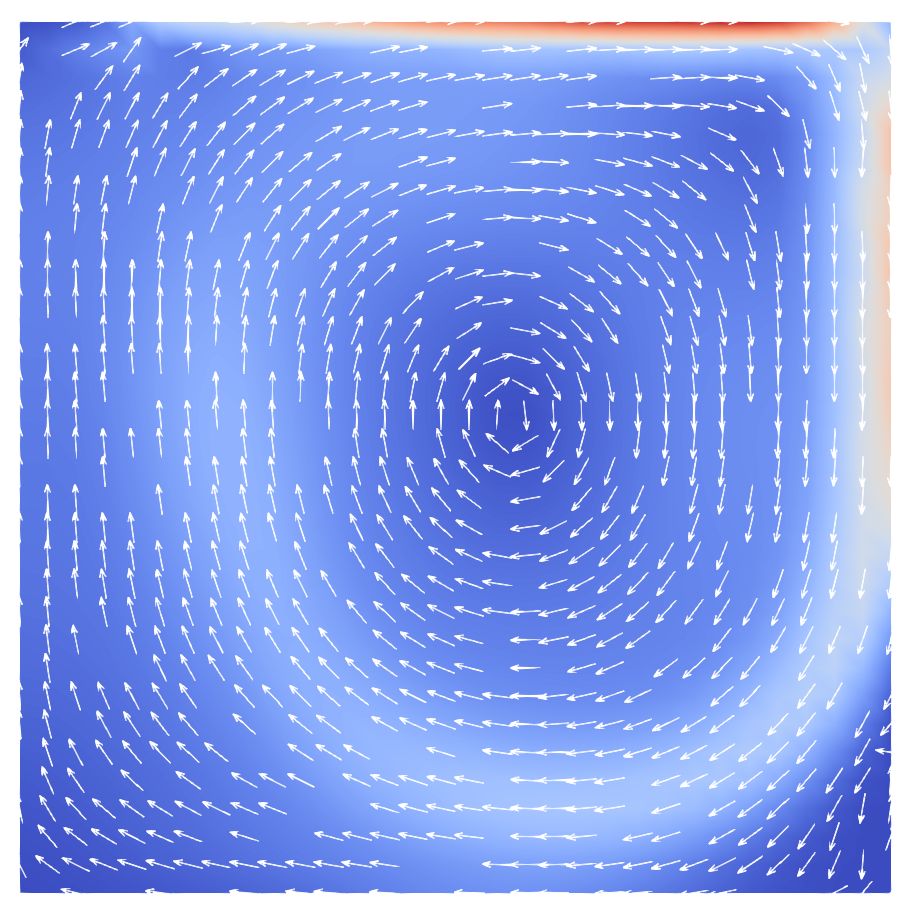}
    \end{subfigure}
    \begin{subfigure}{0.24\linewidth}
        \centering
        \includegraphics[width=\linewidth]{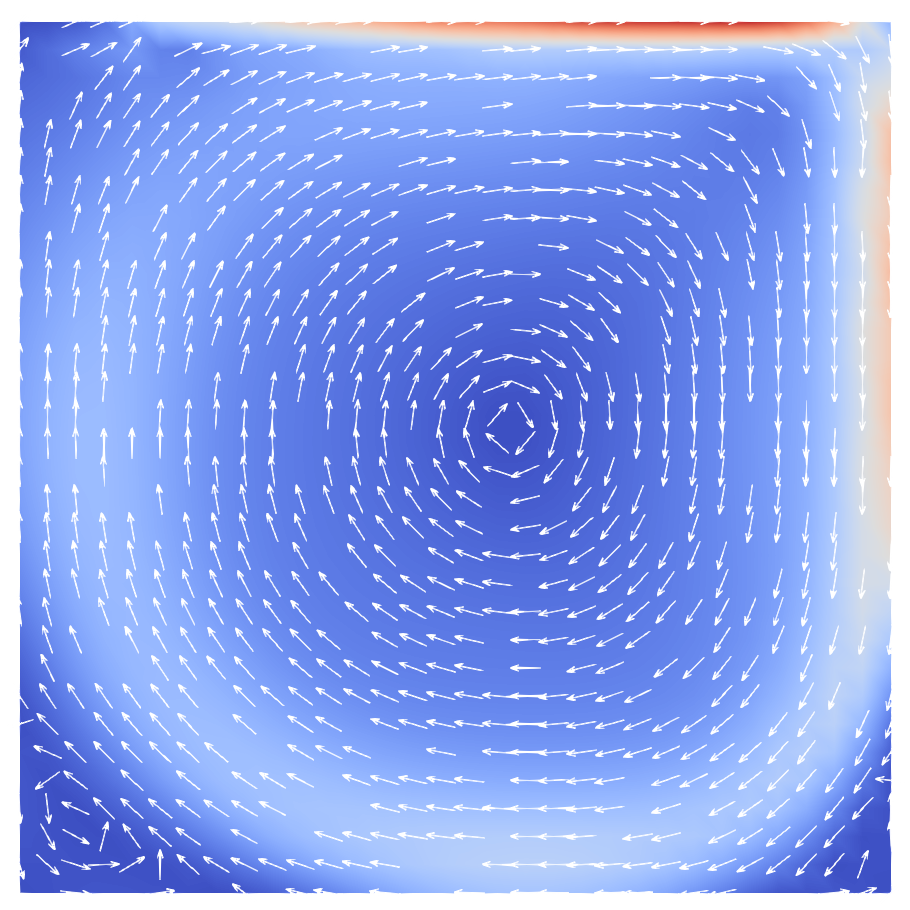}
    \end{subfigure}
    
    \begin{subfigure}{0.24\linewidth}
        \centering
        \includegraphics[width=\linewidth]{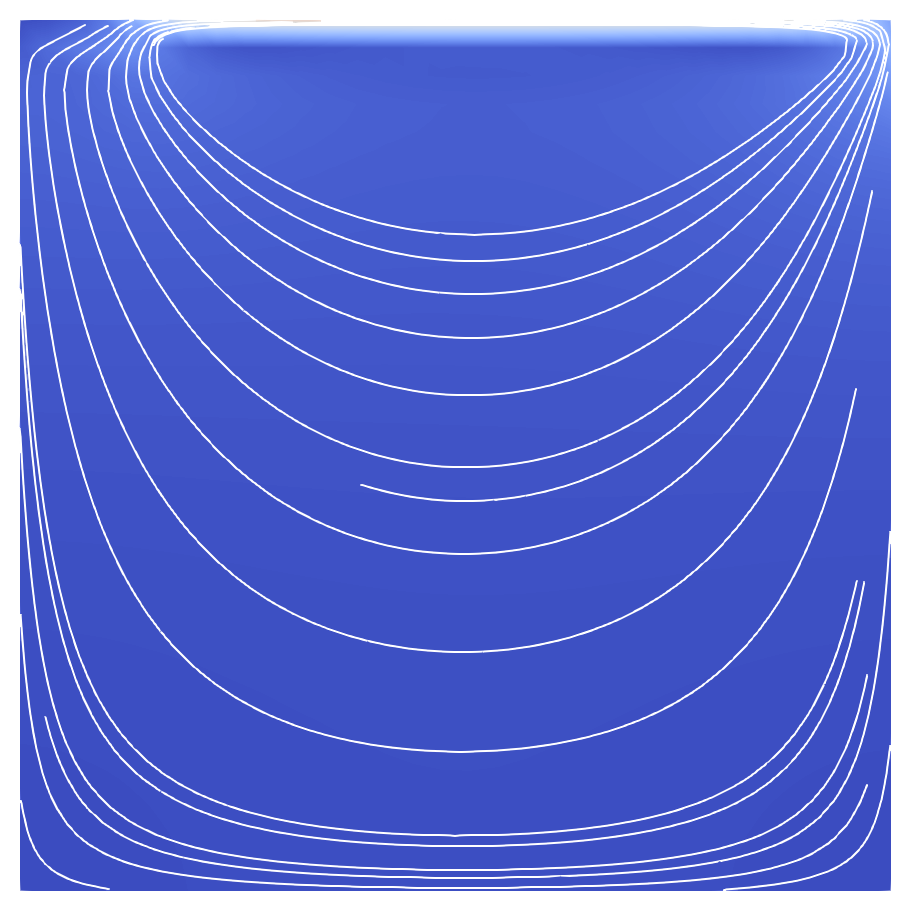}
         \caption{$t = 0.4s$}
    \end{subfigure}
    \begin{subfigure}{0.24\linewidth}
        \centering
        \includegraphics[width=\linewidth]{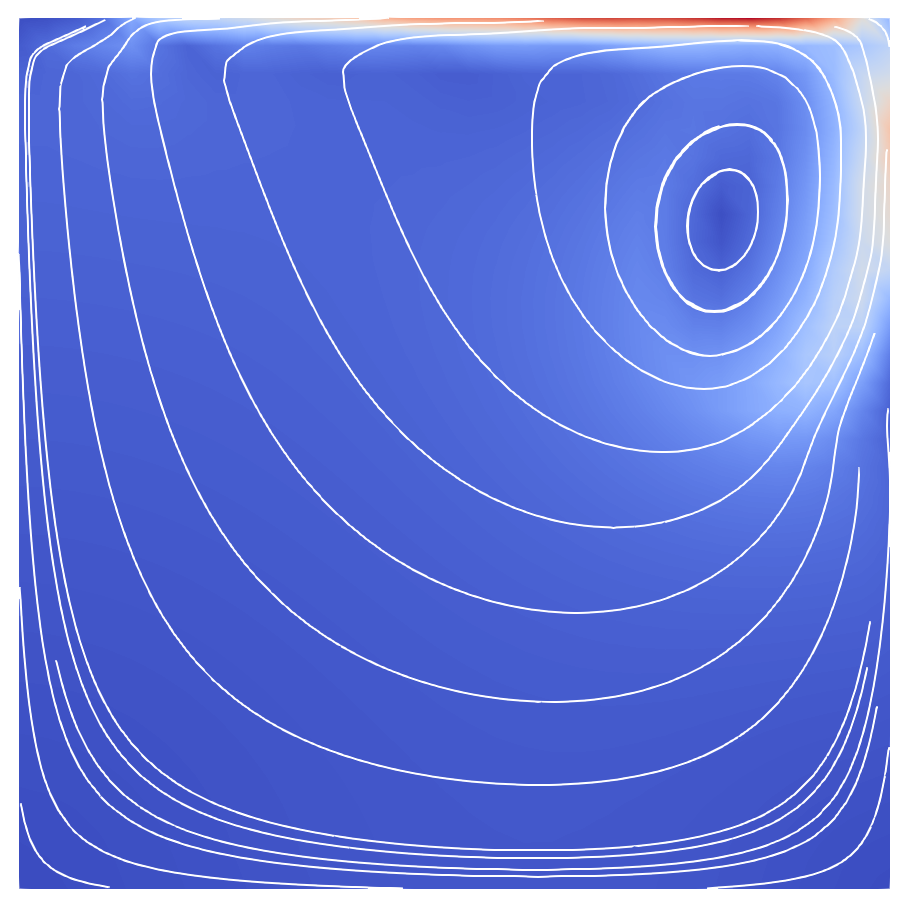}
        \caption{$t = 4s$} 
    \end{subfigure}
    \begin{subfigure}{0.24\linewidth}
        \centering
        \includegraphics[width=\linewidth]{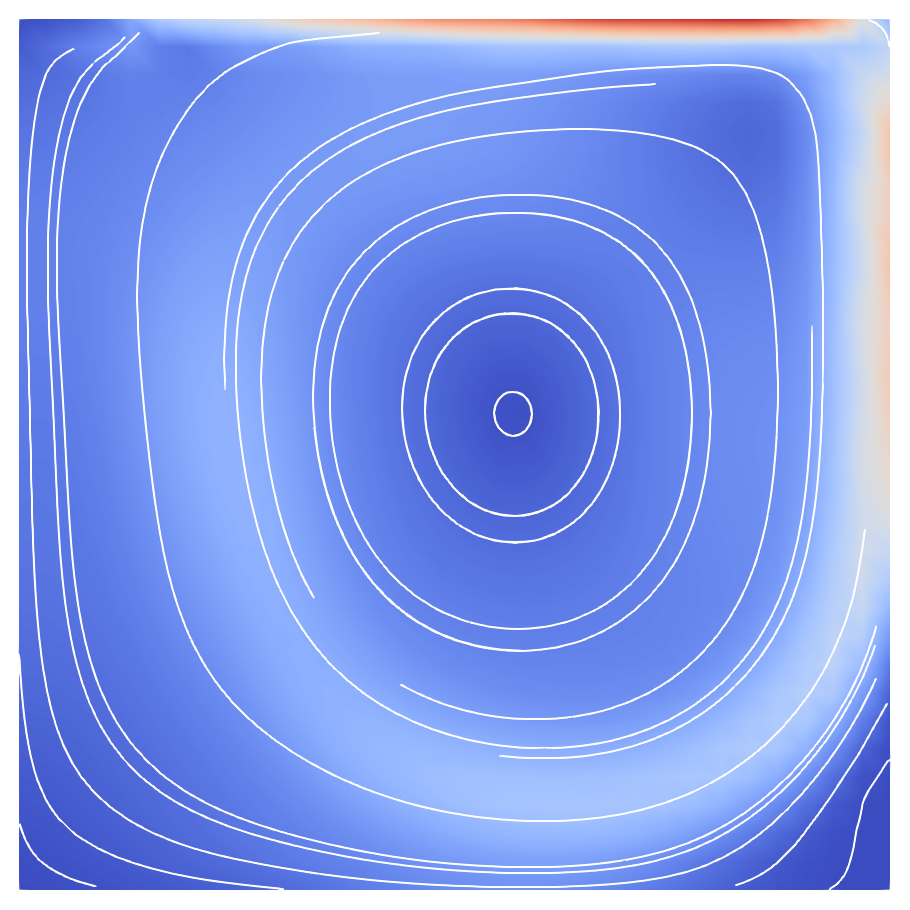}
        \caption{$t = 18s$}
    \end{subfigure}
    \begin{subfigure}{0.24\linewidth}
        \centering
        \includegraphics[width=\linewidth]{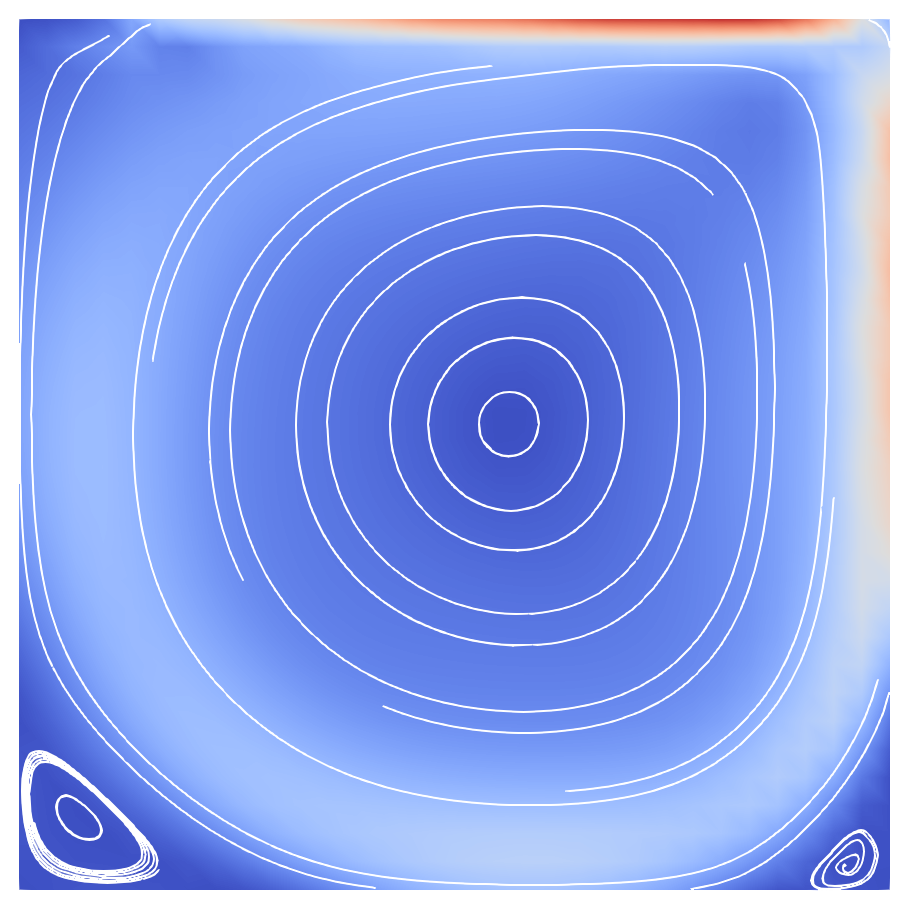}  
        \caption{$t = 28s$}
    \end{subfigure}
    \end{minipage}
    \begin{minipage}{.059\linewidth}
        
        \centering
        \includegraphics[width=\linewidth]{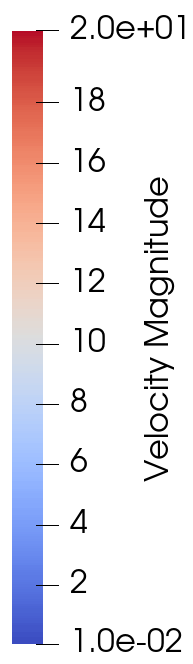}
    
    \end{minipage}

    
    \caption{Visualization of the velocity field at different time steps, sliced along the x-y plane. The top row of figures displays the direction of flow at each point while the bottom row shows streamline plots. }
	\label{fig:visualization}
\end{figure*}

Having been presented with the challenge of finding a suitable teaching code for our two-semester course series \enquote{Performance Engineering}, we developed \thename\footnote{\thename is \acronymexplanation}, a computational fluid dynamics (CFD) proxy app designed for teaching performance analysis and optimization.
\thename solves a variant of the lid-driven cavity problem, an established CFD benchmark problem.
It simulates the flow of an incompressible medium confined in a cubic volume where one of the walls introduces tangential motion~\cite{kuhlmann_lid-driven_2019}, causing a vortex inside the cavity.

To make \thename suitable for use in teaching, we designed it to meet the specified requirements. Solving fluid dynamics using sparse linear algebra is a common workload in HPC, aligning our implementation with many real-world applications.
This ensures that students can transfer learning outcomes to future projects.  
\thename is implemented as a small application, available under a permissive open-source license, comprising fewer than 1,100 lines of code.
This compact design minimizes the time required for students to become familiar with the code.
All performance-relevant data structures and algorithms are implemented in a self-contained manner in C++, avoiding reliance on external libraries or frameworks.
This design ensures that relevant computations are not hidden in library calls, providing students with complete control over the code’s performance.
Finally, while the code is parallelized in a straightforward manner using OpenMP, performance analysis reveals significant opportunities for optimization in both single-core performance and the parallelization scheme.  



The remainder of this paper is structured as follows:
In \autoref{sec:related-work}, we give an overview of relevant previous work.
The details of the simulated problem, our computational approach and the implementation design are described in \autoref{sec:impl}.
This section also discusses possible performance optimizations that could be performed as part of student assignments.
We elaborate on the concept of our course series \enquote{Performance Engineering}, in which we use \thename in teaching, in \autoref{sec:course-design}, before outlining future work and concluding the paper in \autoref{sec:conclusion-future-work}.

\section{Related Work}
\label{sec:related-work}
Although high-performance computing (HPC) is a highly active research field where performance and efficiency play a crucial role, literature on how to teach the fundamentals of performance analysis and optimization is scarce.
While previous work exists that documents experiences and concepts of teaching HPC~\cite{joiner_teaching_2006, neeman_supercomputing_2008, fitz_gibbon_teaching_2010, moore_flipped_2016, chen_what_2021} and educators have argued to prioritize performance aspects in computer science education~\cite{dugan_performance_2004}, we are aware of only a few other projects developing a code specifically designed for teaching performance engineering.

The work most similar to ours is \textit{miniWeather}, a code simulating fluid flows inspired by weather simulation~\cite{miniWeather}.
It is specifically designed for teaching, but focuses more on how to use different parallel programming models instead of low-level performance optimization.
\textit{MD-Bench} is a molecular-dynamics simulation code, designed as an accessible testbench for performance prototyping and algorithm evaluation~\cite{ravedutti_lucio_machado_md-bench_2023}.
While it specifically focuses on performance research and education, its use cases cover more than teaching alone, resulting in a codebase approximately ten times larger than that of \thename.
\textit{CFDPython} is an open-source collection of Python notebooks, structured in lessons and designed to teach students the fundamentals of CFD simulation.
Instead of performance, it focuses on the computational foundations, numerical methods and the underlying physics~\cite{cfd_python}.

Outside of teaching, using proxy apps that capture the performance characteristics of full HPC applications while remaining more manageable with fewer lines of code and dependencies is a well-established practice in HPC research.
The U.S. Exascale Computing Project maintained a list of proxy apps that it deemed to be relevant for HPC research~\cite{ecp_proxies}.
The entries were selected to cover the bandwidth of application domains, programming models, and computational methods.
Even though not designed with teaching in mind, some entries could still be useful for educational purposes.

\section{\thename}
\label{sec:impl}
\newcommand{\veloc}{\pmb{\mathrm{v}}}
\newcommand{\press}{\mathrm{p}}
\newcommand{\grav}{\pmb{\mathrm{g}}}

\thename simulates a \textit{lid-driven cavity flow}.
This established benchmark problem describes a cubic enclosed space filled with an incompressible fluid that is being accelerated at one of the walls, leading to the forming of a vortex in the cavity.
To model the state of the fluid inside the cavity using the velocity field $\veloc$ and pressure field $\press$, we approximate the fluid dynamics with Euler's equations, a simplification of the Navier-Stokes equations that neglects viscosity:
\begin{align}
	\frac{\partial \veloc}{\partial t} &= - (\veloc \cdot \nabla) \veloc - \frac{1}{\rho}\nabla\press + \grav \label{euler1} \\
	\nabla \cdot \veloc &= 0 \label{euler2}
\end{align}
$\rho$ is the density of the medium and $\grav$ denotes external forces (e.g., gravity) that act on the fluid.
To model the behavior at the edges of the simulation domain, we use simple Dirichlet boundary conditions. 
A detailed description on the modeling of the fluid dynamics can be found in~\cite{bridson_fluid_2015}.
For \thename, we use~$\grav$ to model the force applied by the \enquote{moving lid}.
To eliminate the need for input processing, the starting conditions of the simulation are hard-coded and describe a stationary medium under equal pressure:
\begin{equation}
	\veloc_0 = 0, \; \press_0 = 1
\end{equation}

\subsection{Computational Approach}
To compute an approximate numerical solution to the problem described by the equations above, the system is discretized in both space and time.
The velocity and pressure fields are represented using a \textit{staggered grid}, a discretization approach in which velocity components are stored at cell borders.
This configuration simplifies the computation of divergence~\cite{harlow_staggered_grids}.
Temporal discretization is achieved using fixed-length time steps.
During each time step, the simulation state is updated using the \textit{fractional step method}~\cite{stam_stable_fluids}, which divides the computation into sequentially evaluated substeps.

At the start of each time step, external forces $\grav$ are incorporated by adding a time-invariant unidirectional offset to the velocity field.
Subsequently, the advection of the velocity field is computed using a semi-Lagrangian scheme. Finally, to enforce the incompressibility condition, a pressure correction is applied.
This involves computing the velocity field's divergence, solving a Poisson equation for the pressure, and applying the resulting correction to the velocity field.

The Poisson equation results in a system of linear equations that we represent in memory as a sparse matrix using \textit{Compressed-Row Storage} (CSR) and then solve it using a \textit{Preconditioned Conjugate Gradient} (PCG) solver.
To exploit the sparse matrix representation, the matrix-vector product is implemented with respect to the representation (SpMV).
Depending on the simulation configuration, we employ either a \textit{Simplified Diagonal-Based Incomplete Cholesky} (DIC) or a \textit{Jacobi} preconditioner to improve the matrix condition and lower the number of PCG iterations.


\newcommand{\cs}[1]{\Verb"#1"}
\renewcommand{\theadfont}{\normalsize\bfseries}

\begin{table*}[t]
    \centering
    \rowcolors{2}{white}{gray!15}
    \begin{tabular}{l||r|r|r||r|r|r||r|r}
        \toprule
        & \multicolumn{3}{c||}{\thead{Unoptimized}} & \multicolumn{3}{c||}{\thead{Optimized}} & \multicolumn{2}{c}{\thead{Change}}\\
        \rowcolor{white}
        \thead{Code section} & \thead{Calls} & \thead{Incl.\\rt.\\{[s]}} & \thead{Rt.\\frac.\\{[\%]}} & \thead{Calls} & \thead{Incl.\\rt.\\{[s]}} & \thead{Rt.\\frac.\\{[\%]}} & \thead{Ttl.\\{[\%]}} & \thead{Per\\call\\{[\%]}}\\
        \midrule
\cs{cfdSCOPE} & \num{1} & \num{98.82} & \num{100.00} & \num{1} & \num{24.00} & \num{100.00} & \num{-75.71} & \num{-75.71}\\
\cs{├─ applyForces} & \num{15} & \num{0.24} & \num{0.25} & \num{15} & \num{0.25} & \num{1.05} & \num{3.72} & \num{3.72}\\
\cs{├─ solveAdvection} & \num{15} & \num{9.43} & \num{9.54} & \num{15} & \num{9.19} & \num{38.30} & \num{-2.51} & \num{-2.51}\\
\cs{├─ solvePressureCorrection} & \num{15} & \num{75.68} & \num{76.58} & \num{15} & \num{12.87} & \num{53.64} & \num{-82.99} & \num{-82.99}\\
\cs{│\ \ └─ pcg} & \num{15} & \num{70.47} & \num{71.31} & \num{15} & \num{12.37} & \num{51.53} & \num{-82.45} & \num{-82.45}\\
\cs{│\ \ \ \ \ ├─ precondition} & \num{1924} & \num{34.63} & \num{35.04} & \num{5821} & \num{0.82} & \num{3.41} & \num{-97.63} & \num{-99.22}\\
\cs{│\ \ \ \ \ ├─ spmv} & \num{1924} & \num{5.68} & \num{5.75} & \num{5821} & \num{7.26} & \num{30.25} & \num{27.75} & \num{-57.78}\\
\cs{│\ \ \ \ \ ├─ dot} & \num{7666} & \num{7.42} & \num{7.51} & \num{23269} & \num{2.54} & \num{10.59} & \num{-65.76} & \num{-88.72}\\
\cs{│\ \ \ \ \ ├─ operator+} & \num{5742} & \num{8.05} & \num{8.15} & \num{15} & \num{0.05} & \num{0.22} & \num{-99.34} & \num{154.46}\\
\cs{│\ \ \ \ \ ├─ operator*} & \num{5727} & \num{7.93} & \num{8.02} & - & - & - & - & -\\
\cs{│\ \ \ \ \ └─ multiply\_add\_inplace} & - & - & - & \num{17418} & \num{1.54} & \num{6.43} & - & -\\
\cs{├─ applyPressureCorrection} & \num{15} & \num{0.08} & \num{0.08} & \num{15} & \num{0.08} & \num{0.34} & \num{8.81} & \num{8.81}\\
\cs{└─ write\_to\_file} & \num{16} & \num{13.06} & \num{13.21} & \num{16} & \num{1.35} & \num{5.61} & \num{-89.69} & \num{-89.69}\\
        \bottomrule
    \end{tabular}
    \caption{Computational hotspots of both \thename variants for the default problem size (see \autoref{sec:impl}), showing call counts (\enquote{Calls}), inclusive runtimes (\enquote{Incl. rt.}), and their proportion of overall runtime (\enquote{Rt. frac.}). The relative differences in total runtime (\enquote{Ttl.}) and per-call runtime between the unoptimized and optimized versions are also provided. Functions absent in one of the variants are indicated with \enquote{-}.}
    \label{tab:hotspots}
\end{table*}


\subsection{Implementation}
The \thename implementation does not rely on dependencies for the computation itself and all data structures and algorithms have been implemented from scratch, including the sparse matrix representation, the basic matrix operations (e.g., SpMV) and the \textit{Preconditioned Conjugate Gradient} (PCG) linear algebra solver.
Two external libraries, \textit{spdlog}\footnote{\url{https://github.com/gabime/spdlog}} for formatted logging and \textit{cxxopts}\footnote{\url{https://github.com/jarro2783/cxxopts}} for command-line-argument parsing, are used for setup code that is irrelevant for performance.
Both libraries are downloaded automatically by CMake when configuring the project and are built together with the project to maintain a straightforward build process.

To allow for visual inspection of the simulation output, the application can be configured to save the simulation state on the file system after each time step.
The output files use a simple CSV format and contain the current states of the velocity and pressure fields.
Along with the simulation code, we also provide a script to parse and visualize the output files using \textit{ParaView}\footnote{\url{https://www.paraview.org}}.
An example visualization, showing the formation of the vortex in the simulated medium, is depicted in \autoref{fig:visualization}.

We employ OpenMP for multi-threaded computation of all functions that allow for parallel execution.
These include the basic vector operations (addition, multiplication), the sparse-matrix-vector multiplication (SpMV) and the implementation of the semi-Lagrangian advection of the velocity field.
The parallelization in \thename is intentionally kept simple to resemble codes with retrofitted parallelization and projects where parallelism was not part of the initial design, which are often the focus of performance engineering.
Currently, \thename does not support distributed-memory parallelism, such as MPI.

The source code includes a suite of unit tests to ensure the correctness of key components, such as fundamental linear algebra functions and basic simulation routines.
Additionally, an integration test is provided, which runs a small-scale simulation and compares its output to a reference snapshot.
This setup helps students check that their optimizations do not inadvertently affect the accuracy of the computations, though it cannot guarantee the detection of all bugs.
The tests can be deactivated during configuration and are compiled into separate binaries, inhibiting any influence on the performance of the simulation itself.

\subsection{Optimization Opportunities}
\thename is designed as an educational example code for teaching performance engineering concepts.
To demonstrate its suitability for this purpose, we also developed an optimized version that highlights its potential for runtime improvements.
Beginning with the unoptimized version, we followed the typical performance engineering workflow, which students would also employ.
This process first analyzes the code’s performance characteristics and then iteratively improves its performance through a cycle of code modifications and performance measurements.
All measurements for this paper have been conducted on a dedicated compute-node of the \lbcluster, featuring two Intel Xeon Platinum 9242 processors with a total of 104 cores and \SI{512}{\giga \byte} of main memory.
The CPU was set to a fixed frequency of \SI{2101}{\mega \hertz}.
Unless stated otherwise, all measurements refer to a \thename run using \num{32} OpenMP threads over \SI{6}{\second} simulation time, with a step size of \SI{0.4}{\second}, a simulation domain with side length of~\num{100} and all other parameters left to their default values.
To generate profile and trace data about the simulation runs, we employed the \textit{Score-P} measurement infrastructure~\cite{Knuepfer2012} and used \textit{PAPI}\footnote{\url{https://github.com/icl-utk-edu/papi}} to collect hardware performance counters.

\begin{figure*}
	\centering
    \includegraphics{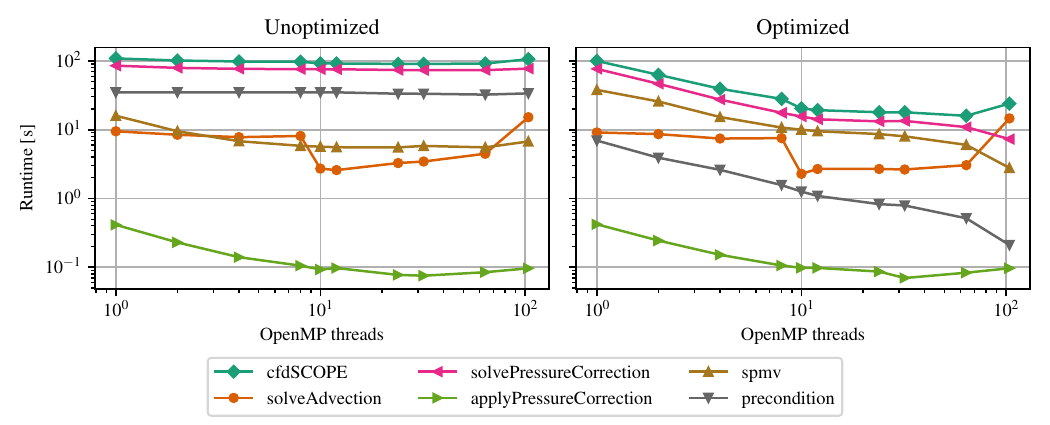}
	\caption{Strong scaling plot for the unoptimized and optimized versions of \thename, depicting the runtime of the simulation and some a selection if its functions for different numbers of OpenMP threads. Both axes are logarithmic.}
	\label{fig:strong-scaling}
\end{figure*}

\begin{figure}
	\centering
	\includegraphics{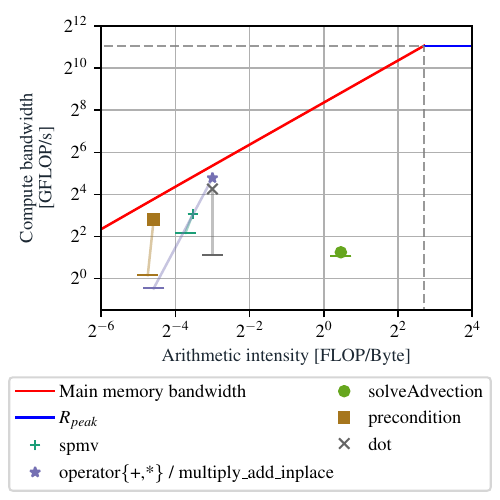}
	\caption{Roofline model for selected kernels. The sloped red and horizontal blue lines depict the achievable performance as a function of a kernel's arithmetic intensity, consisting of the memory bandwidth and maximum compute bandwidth $R_{peak}$, respectively. Limits have been measured using 32 CPU cores. Both axes are logarithmic. For each kernel, the horizontal dash describes the performance of the unoptimized version, while the connected marker represents the optimized version. The entry with the star marker shows the performance of vector addition/multiplication (\texttt{operator+}, \texttt{operator*}) in the unoptimized version and \texttt{multiply\_add\_inplace} in the optimized version.}
	\label{fig:roofline}
\end{figure}

A useful first step in analyzing a code's performance are overview measurements to identify computational hotspots.
This allows a performance engineer to focus effort on the sections of the application that account for a significant portion of the overall runtime.
\autoref{tab:hotspots} lists the hotspots in \thename, along with their respective runtimes for both the unoptimized and optimized versions.
The measurements show that our optimized version reduces the simulation runtime by \SI{76}{\percent}, with the majority of the improvement located in the computation of the pressure correction, where the runtime was reduced by \SI{83}{\percent}.

To evaluate the parallel efficiency of a code, performance analysts commonly perform scaling measurements. 
\autoref{fig:strong-scaling} illustrates the strong scaling behavior, showing how runtime varies with increasing thread counts for a fixed problem size in both code versions.
The results indicate that while certain sections of the code experience runtime reductions with increased parallelism, the overall runtime of the unoptimized version shows minimal improvement.
In fact, using more than \num{30} OpenMP threads appears to degrade performance.
In contrast to that, the optimized version exhibits significantly improved strong scaling behavior up to approximately 32 threads.

Overall, comparing the measurements between the unoptimized and optimized version demonstrates the considerable potential for improvement of the unoptimized version.
Hence, the code provides students with ample opportunities to experiment with various performance optimization techniques and achieve significant speedups.
In the remainder of this section, we explore some optimization opportunities available in the code by explaining the key changes implemented in our optimized version:

\paragraph{Algorithmic changes}
In many cases, identifying and eliminating redundant or unnecessary computations can lead to significant performance improvements, and these optimizations can often be implemented without detailed knowledge of the underlying hardware architecture.
In our case, the system of linear equations solved during the pressure correction is represented by a symmetric matrix.
By leveraging this symmetry, the memory required for the assembled matrix is reduced by storing only the main diagonal and one triangle of the matrix, thereby improving cache efficiency.
Further optimization is possible by exploiting the matrix structure more extensively, following approaches similar to those outlined in~\cite{burger_memory-ecient_2017}.
Together, these optimization lead to the improved performance and scaling behavior of the SpMV product that is visible in \autoref{fig:roofline} and \autoref{fig:strong-scaling}.
Additionally, although the matrix entries change between simulation time steps, the sparsity pattern, i.e., the locations of non-zero entries, remains constant across all time steps and can be reused.
This significantly reduces the computational effort required to prepare the matrix at each time step.

\paragraph{Matrix preconditioning}
The measurements in \autoref{tab:hotspots} indicate that the matrix preconditioner consumes a significant portion of the overall runtime.
In the unoptimized version, a \textit{Diagonal-based Incomplete Cholesky} (DIC) preconditioner is used. While this preconditioner effectively reduces the number of PCG iterations, it accounts for a considerable fraction of the runtime, as shown in \autoref{tab:hotspots}.
Furthermore, its computation cannot be easily parallelized due to dependencies between loop iterations, causing its runtime to remain constant regardless of the number of OpenMP threads. 

To address this, the optimized version employs a simpler Jacobian preconditioner.
Although it does not reduce the number of PCG iterations as effectively, it is fully parallelized, allowing its performance to scale with increased thread counts, c.f.,~\autoref{fig:strong-scaling}.
Combined with other optimizations that reduce the computational cost of the PCG solver, this change proves to be beneficial overall.

\paragraph{Reducing memory allocations}
A detailed review of the unoptimized source code reveals several instances where unnecessary memory allocations and copies occur.
This overhead can be eliminated by modifying operations such as vector addition and multiplication, the matrix preconditioner, and sparse matrix-vector multiplication to work in-place on pre-allocated operands.
Additionally, the optimized version introduces the \texttt{multiply\_add\_inplace} function, which combines vector multiplication and addition into a single operation.
This enables the compiler to generate \textit{fused multiply-add} (FMA) vector instructions and increases the arithmetic intensity.
\autoref{fig:roofline} illustrates the significant effect that these changes have on the compute bandwidth.

\paragraph{Exposing more parallelism}
The unoptimized version includes several functions that can benefit from thread-level parallelism (TLP) without requiring algorithmic changes.
These functions include vector operations, such as the dot product, which are currently parallelized only at the instruction level (ILP), and the matrix assembly for the pressure correction solver.
Modifying these functions to use simple OpenMP worksharing loops significantly increases their compute bandwidth and enhances their scaling performance, as shown in \autoref{fig:roofline} and \autoref{fig:strong-scaling} respectively.

\paragraph{Improving memory access patterns}
The unoptimized version contains several sections with memory access patterns that lead to suboptimal cache usage and, in multi-threaded execution, to memory contention.
These have been revised in the optimized version, for example by introducing a blocking iteration scheme in the computation of the advection.
Moreover, tuning of the OpenMP parameters (e.g., the \texttt{collapse} and \texttt{schedule} clauses for worksharing loops) and the granularity of the parallelization proved to be useful to further increase performance.

\paragraph{I/O optimizations}
In the unoptimized implementation, the simulation state is written to a file after each time step using a single thread.
Switching to a multi-threaded approach, where threads write to local buffers that are merged and saved as a block, considerably enhances I/O performance, as apparent in \autoref{tab:hotspots}.
This improvement is also evident in the traces of the optimized version in \autoref{fig:traces}, which depicts multiple threads operating during the I/O routine, now moved to the beginning of the timestep.

\captionsetup[subfigure]{aboveskip=-8pt,belowskip=+0pt}

\begin{figure*}
	\centering
    \begin{subfigure}[b]{0.98\textwidth}
	   \includegraphics[width=\textwidth]{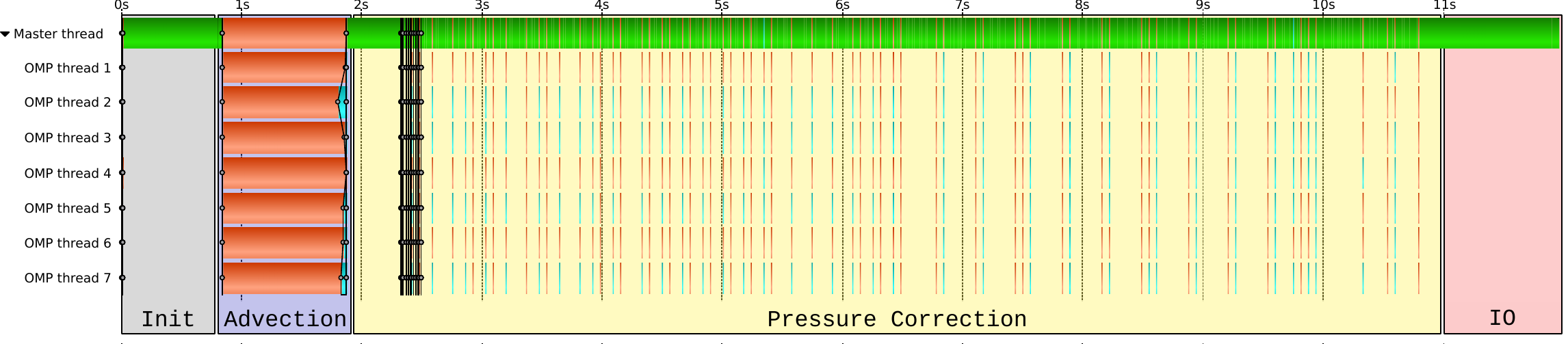}  
      \label{subfig:trace-before}
      \caption{Trace before optimization.}
    \end{subfigure}
    \par\medskip
    \begin{subfigure}[b]{0.98\textwidth}
      \includegraphics[width=\textwidth]{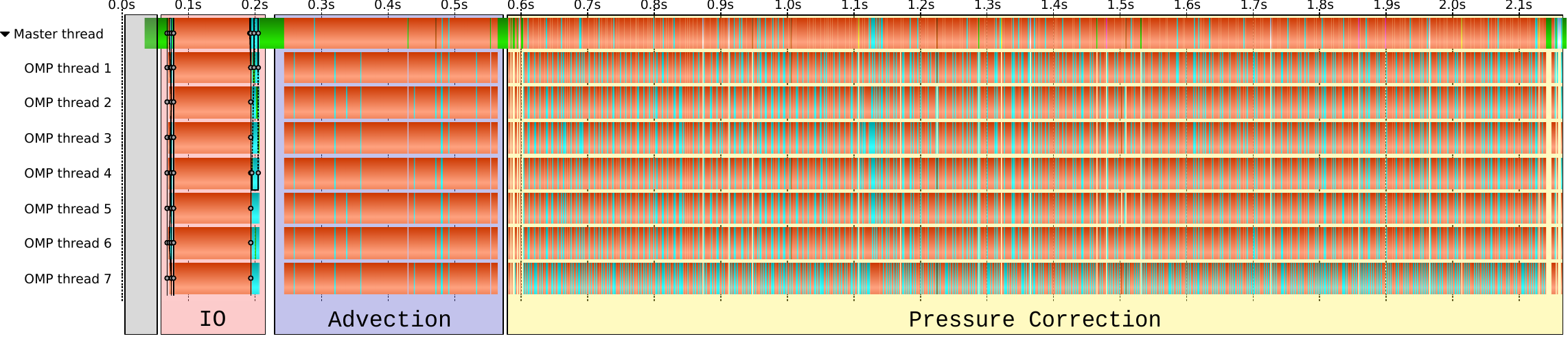}
      \label{subfig:trace-after}
      \caption{Trace after optimization.}
    \end{subfigure}
    \caption{Comparison of traces before and after optimization. Shown is a single time step computed with 8 threads. Note that the time scales differ between the two traces.}
    \label{fig:traces}
\end{figure*}

\section{Course Design}
\label{sec:course-design}
We use \thename in academic teaching as part of our two-semester course series \enquote{Performance Engineering} (PEng).
This course introduces graduate and advanced undergraduate students of computer science and related fields to the fundamentals of performance analysis and optimization in an HPC context.  
The series consists of two distinct courses that we recommend students to take in sequence.
The first part is a seminar focused on performance analysis.
After four introductory lectures covering the basics of benchmarking, performance analysis tools, and methods for understanding the performance behavior of HPC software, students analyze the unoptimized version of \thename from a performance perspective. They document their findings in a brief seminar report.  
Students are free to choose their tools and methods for the analysis, though we provide recommendations and guidance if needed. 
At the end of the semester, each student delivers a brief 
talk on their results.
We encourage peer questions after the presentations, ideally fostering discussions about the methods and results.

The second part of the series is a practical course titled \enquote{Performance Engineering: Hands-On}, focusing on performance optimization.
Following three lectures and two short assignments on relevant topics such as blocking, vectorization and compiler optimizations, students are tasked with applying the insights gained in the seminar to improve \thename's runtime performance.
As a starting point, all participants are provided with the seminar reports from the previous course on performance analysis, which point to the performance bottlenecks as promising targets for optimizations.  
Similar to the seminar, students document their optimization efforts in a brief report and give a short presentation.
No strict guidelines are imposed on how they approach the task, allowing for creativity and experimentation.  

To introduce an element of competition, we host a \enquote{performance challenge}.
Students benchmark their best optimization results in a defined benchmarking environment on our HPC cluster and are encouraged to post their runtimes on a course leaderboard.
At the end of the semester, the student with the fastest runtime is awarded a prize.
While the challenge fosters competitive engagement, participation is voluntary, and a student’s performance relative to their peers is not relevant for the course grade.

\section{Conclusion and Future Work}
\label{sec:conclusion-future-work}
We introduced \thename, a computational fluid dynamics (CFD) proxy app designed to be used in teaching performance analysis and optimization.
The application implements a simulation of the lid-driven cavity problem, an established benchmark problem CFD.
All performance-relevant parts have been implemented from the ground up to give students full control over the performance characteristics.
To demonstrate that the unoptimized version exhibits substantial potential for optimization, and thus learning opportunities, we also presented an optimized version that improves the runtime performance by \SI{76}{\percent}.

Currently, \thename's primary limitation is its support exclusively for shared-memory parallelism, with no capability for distributed memory architectures.
To overcome this, future work could incorporate a mesh-partitioning mechanism, enabling multi-node simulation runs.
Adapting the simulation to support massively parallel or hybrid computing architectures also presents a promising opportunity for further development, aligning with recent trends in HPC.

The optimized version, intended as a lower bound for potential runtime improvements, is unlikely to represent the performance optimum.
We anticipate that future optimization efforts could yield even greater performance gains.
For instance, these efforts might focus on further optimizing the advection kernel or experimenting with other matrix preconditioners in the PCG.
Additionally, future work could explore implementing more advanced physics (e.g., viscosity or varied boundary conditions) or incorporating more complex geometries.

\thename is licensed under the MIT license.
The unoptimized version is available on GitHub: \repo.
The optimized version will be shared with HPC researchers and educators upon request to the authors.

\ifx\myanonymous\undefined
\begin{acks}
The authors sincerely thank Yannic Fischler for his outstanding support in software and dependency management, which greatly facilitated the performance measurements for this work.
They also gratefully acknowledge the computing time provided to them at the NHR Center NHR4CES at TU Darmstadt.
This is funded by the Federal Ministry of Education and Research, and the state governments participating on the basis of the resolutions of the GWK for national high performance computing at universities.
\end{acks}
\fi

\bibliographystyle{ACM-Reference-Format}
\bibliography{bibliography}


\fi

\end{document}